\documentclass[aps,twocolumn,showpacs,preprintnumbers,amsmath,amssymb]{revtex4}
\usepackage{epsf}
\usepackage{graphicx} 
\usepackage{ulem} 
\usepackage[usenames]{color}
\newcommand{\ba}{\begin{eqnarray}}
\newcommand{\ea}{\end{eqnarray}}
\begin{document}
\title{The role of the $N^*(1535)$ resonance and the $\pi^-p\to KY$ amplitudes in the OZI forbidden $\pi N \to
\phi N$ reaction}
\author{M. D\"oring$^1$, E. Oset$^2$ and B.S. Zou$^{3,2}$}
\affiliation{
$^1$ Institut f\"ur Kernphysik, Forschungszentrum J\"ulich GmbH, 52425 J\"ulich, Germany\\
$^2$ Departamento de F\'{\i}sica Te\'orica and IFIC,
Centro Mixto Universidad de Valencia-CSIC,
Institutos de Investigaci\'on de Paterna, Aptdo. 22085, 46071 Valencia, Spain\\
$^3$Institute of High Energy Physics, CAS, Beijing 100049, China
}
\pacs{
14.20.Gk, 
12.39.Fe, 
13.60.Le, 
14.40.Cs  
}

\begin{abstract} 
We study the $\pi N \to \phi N$ reaction
close to the $\phi N$ threshold within the chiral unitary approach, by 
combining the 
$\pi^- p \to K^+ \Sigma ^-$, $\pi^- p \to K^0 \Sigma ^0$ and
$\pi^- p \to K^0 \Lambda$ amplitudes with the coupling of the $\phi$ to 
the $K$
components of the final states of these reactions via quantum loops. We 
obtain a
good agreement with experiment when the dominant $\pi^- p \to K^0 \Lambda$
amplitude is constrained with its experimental cross section. We also 
evaluate
the coupling of the $N^*(1535)$ to $\phi N$ and find a moderate coupling 
as a
consequence of partial cancellation of the large $KY$ components of the
$N^*(1535)$. We also show that the 
$N^*(1535)$ pole approximation is too small to reproduce the measured cross
section  for the $\pi^- N \to \phi N$ reaction.
\end{abstract}

\maketitle

\section{Introduction}

  The $\pi^- p \to \phi n$ reaction, assuming the $\phi$ a pure $s \bar{s}$
state, is OZI forbidden and, as a consequence, it should have a very small cross
section compared to analogous OZI allowed ones. Actually, the 
$\pi^- p \to \phi n$ cross section close to threshold \cite{Landolt} is about a
factor fifty smaller than that of the $\pi^- p \to \omega n$ reaction 
\cite{Miller:1969ca,Holloway:1974cp,Lykasov:1998ma}. Yet, that cross section 
is still less suppressed than one might expect.  The usual way reactions escape
the OZI restrictions is through intermediate steps (like loops) which make the
initial state couple to some state with both $u,d$ and $s$ quarks, which in a
second step couples to the pure  $s \bar{s}$ of the $\phi$. This is the case for
instance for the $\phi \to \pi^0 \pi^0 \gamma$ reaction 
\cite{Achasov:2000ym,Aloisio:2002bt}, which is reproduced fairly well in terms of
kaon loops where the kaons couple to the $\phi $ on one side and to the
nonstrange components on the other side 
\cite{Marco:1999df,Markushin:2000fa,Palomar:2003rb}.

   In \cite{Xie:2007qt} it was suggested that the agent responsible for the
relatively large $\pi^- p \to \phi n$ cross section was the $N^*(1535)$
resonance which should have a large coupling to the $\phi N$ system.  The
reasoning behind this suggestion was the large coupling of the $N^*(1535)$ to
meson baryon states with strangeness, in simpler words the large $s \bar{s}$
content of the $N^*(1535)$. Indeed, the analysis of the 
$J/\psi \to \bar{p}\Lambda K^+$ decay and the
$pp \to p\Lambda K^+$ reaction near threshold concluded that the
$N^*(1535)$ resonance has a significant coupling to $K \Lambda$ ~\cite{liu}. 
The analyses \cite{Mosel,Saghai} of the recent SAPHIR and CLAS $\gamma
p\to K^+\Lambda$ data \cite{ELSA,Bradford:2005pt} also show a large coupling
of the $N^*(1535)$ to $K\Lambda$. The other reason to support the relevant role
of the $N^*(1535)$ is that the higher energy $S_{11}$ resonance, the $N^*(1650)$
couples very weakly to $\eta N$ and $K \Lambda$ \cite{PDG}.

 From another perspective, chiral unitary theories provide the $N^*(1535)$ as a
 dynamically generated state from the interaction of the octet of the
 pseudoscalar mesons and the octet of stable baryons 
 \cite{Kaiser:1995cy,Nacher:1999vg,Nieves:2001wt,Inoue:2001ip,Hyodo:2002pk} and
 the large coupling of the resonance to the $\eta N$, $K \Lambda$ and 
 $K \Sigma$ stems naturally from the information of the chiral lagrangians 
 \cite{Ecker:1994gg,Bernard:1995dp} used as input. Indeed, the good reproduction
 of the $\gamma p\to K^+\Lambda$ and $\gamma p\to K^+\Sigma$ within the chiral 
 unitary approach, using the complete set of Feynman diagrams
 demanded by gauge invariance \cite{Borasoy:2007ku}, comes to stress the role of 
 the dynamically generated $N^*(1535)$ resonance in these reactions.
 
  The claim of a nature as a dynamically generated resonance for the $N^*(1535)$
can be interpreted as having the resonance largely build up of meson baryon
components, which play the dominant role in reactions taking place at low 
energies.
There are indications that some extra conventional three constituent quark
components are also present in the resonance \cite{Hyodo:2008xr}. These
components  would
show up at large $Q^2$ in the electroproduction helicity form factors where the
picture of the $N^*(1535)$ as purely dynamically generated resonance produces a
too fast fall down \cite{Jido:2007sm}. 

  The purpose of the present paper is to do further research regarding the idea
of  \cite{Xie:2007qt} by using the chiral unitary approach \cite{Inoue:2001ip}
where the $N^*(1535)$ resonance is dynamically generated. Similarly to the
successful approach of \cite{Marco:1999df,Markushin:2000fa,Palomar:2003rb} in
the description of the $\phi \to \pi^0 \pi^0 \gamma$ reaction, here we shall
also have the $N^*(1535)$ coupling to the $K \Lambda$ and  $K \Sigma$ components
which later on will couple to the $\phi n$ system. Technically, this means the
reaction will proceed via loops of $K \Lambda$ and  $K \Sigma$. The work of 
 \cite{Jido:2007sm}, which studies the $N^*(1535) \to \gamma ~N$ transition, is 
 also useful here since making use of the vector meson
 dominance hypothesis we can replace a photon by a $\phi$, with the appropriate
 conversion factors, and this will provide the coupling of the resonance to
 $\phi n$, which in \cite{Xie:2007qt} was extracted from the $\pi^- p \to \phi n$
 data. 
 
 Parallely we shall conduct another study in which we  shall use the full and
 energy dependent amplitude provided by the chiral unitary approach, instead of the couplings from the $N^*$ pole position. The two procedures should be identical
 should the $N^*(1535)$ resonance dominate absolutely the 
$\pi^- p \to \phi n$ reaction. However, the chiral unitary approach, which
provides some resonances appearing as poles of the scattering matrix, not only
provides the poles but simultaneously generates a background which is relevant
as soon as we move away from the resonance region. In this sense, since the  
$N^*(1535)$ resonance is about $400$~MeV below the $\phi n$ threshold, one
should expect some differences between these two approaches and we shall
investigate them.

\section{Formalism}
\label{sec:model}
\subsection{Meson baryon transition}
\label{sec:meba}
In the present model, the $\phi$ meson is produced by the coupling of the vector meson to intermediate loops, which are provided by a unitary meson-baryon amplitude in coupled channels, with the interaction derived from the lowest order chiral Lagrangian. For the meson-baryon amplitude, we follow closely the approach used in Ref. \cite{Inoue:2001ip}. The idea for the present study of $\pi N\to \phi N$ is that the coupled channel approach provides $\pi N\to K\Lambda,\,K\Sigma$ transitions, with final states to which the $\phi$ can couple according to its main decay channel $\phi\to \bar{K}K$. 

We have the coupled channels $K^{+} \Sigma^{-}$, $K^{0} \Sigma^{0}$, $K^{0} \Lambda$, $\pi^{-} p$, $\pi^{0} n$ and $\eta n$  for the net charge zero case, and $\pi^{0} p$, $\pi^{+} n$, $\eta p$, $K^{+} \Sigma^{0}$, $K^{+} \Lambda$, and $K^{0} \Sigma^{+}$ for the net charge $+1$ case. 
The meson baryon scattering amplitude is described in Ref. \cite{Inoue:2001ip} by means of the Bethe-Salpeter equation for meson baryon scattering  given by
\begin{equation}
   T = V + V G T \ .
\end{equation}
Based on the $N/D$ method and the dispersion relation \cite{Oller:2000fj}, this integral scattering equation can be reduced to a simple algebraic equation
\begin{equation}
   T = (1-VG)^{-1}\,V
   \label{bse}
\end{equation}
where $T$ is a matrix in the coupled channels $i,j$ that provides the unitary amplitude. The matrix $V$ is the $s$-wave meson-baryon interaction provided by the lowest order of chiral perturbation theory, which is the Weinberg-Tomozawa interaction,
\begin{eqnarray}
V_{i j} &=&
 - C_{i j} \frac{1}{4 f^2}(2\sqrt{s} - M_{i}-M_{j})
 \nonumber \\ &&  \ \ \ \times
\sqrt{\frac{M_{i}+E}{2M_{i}}}
\sqrt{\frac{M_{j}+E^{\prime}}{2M_{j}}}
\label{eq:ampl2}
\end{eqnarray}
with the channel indices $i,j$, the baryon mass $M$, the 
meson decay constant $f$  and the center of mass energy $\sqrt s$. The 
coefficients $C_{ij}$ are the coupling strengths determined by the SU(3) group 
structure of the channels and are given in \cite{Inoue:2001ip}. 
The diagonal matrix $G$ is a meson baryon loop function given in terms of meson 
and baryon propagators by
\begin{eqnarray}
   G(\sqrt s) &=&  i\int \frac{d^{4} q}{(2\pi)^{4}} \frac{M}{E(\vec q)}
    \frac{1}{q^{0} -E(\vec q) + i\epsilon}
   \nonumber \\ && \ \ \ \ \ \ \times
    \frac{1}{(P-q)^{2} - m^{2} + i \epsilon}
    \label{propnorela}
\end{eqnarray}
with  the total energy $P=(\sqrt s, 0, 0,0)$ in the center of mass frame and the meson mass $m$. For the baryon propagator, we use a nonrelativistic form, slightly different from Eq. (\ref{propnorela}). Details of the nonrelativistic approximation can be found in a recent discussion in Ref. \cite{Jido:2007sm}.
In dimensional regularization,  the loop function in each channel $i$ is given by the following analytic expression:   
\begin{eqnarray}
G_i &=& i \,  \int \frac{d^4 q}{(2 \pi)^4} \,
\frac{2 M_i}{q^2 - M_i^2 + i \epsilon} \, \frac{1}{(P-q)^2 - m_i^2 + i
\epsilon}  \nonumber \\ &=& \frac{2 M_i}{16 \pi^2} \left\{ a_i(\mu) + \ln
\frac{M_i^2}{\mu^2} + \frac{m_i^2-M_i^2 + s}{2s} \ln \frac{m_i^2}{M_i^2} +
\right. \nonumber \\ & &  \phantom{\frac{2 M_i}{16 \pi^2}} +
\frac{\bar q_i}{\sqrt{s}}
\left[
\ln(s-(M_i^2-m_i^2)+2 \bar q_i\sqrt{s})
\right. \nonumber  \\
&&  \phantom{\frac{2 M_i}{16 \pi^2} +\frac{\bar q_i}{\sqrt{s}}}
  \hspace*{-0.6cm}+ \ln(s+(M_i^2-m_i^2)+2 \bar q_i\sqrt{s}) \nonumber  \\
& & \phantom{\frac{2 M_i}{16 \pi^2} +\frac{\bar q_i}{\sqrt{s}}}
  \hspace*{-0.6cm}- \ln(-s+(M_i^2-m_i^2)+2\bar q_i\sqrt{s}) \nonumber  \\
& & \left. \left. 
     \phantom{\frac{2 M_i}{16 \pi^2} +\frac{\bar q_i}{\sqrt{s}}}
 \hspace*{-0.6cm} - \ln(-s-(M_i^2-m_i^2)+2\bar q_i\sqrt{s}) \right]
\right\} ,
\label{eq:gpropdr}
\end{eqnarray}
where $\bar q_i$ is the 3-momentum of the meson or baryon in the center of mass
frame, $\mu$ is the scale of dimensional regularization and $a_i(\mu)$ are
subtraction constants, which are determined by a fit to the $S_{11}$ and
$S_{31}$ partial waves of $\pi N$ scattering \cite{Inoue:2001ip}. There
are four independent subtraction constants in Ref. \cite{Inoue:2001ip}, one for
each intermediate state, taking just one subtraction constant for states
belonging to the the same isospin multiplet. Thus the free parameters of the theory are $a_{\pi N}$, $a_{K\Sigma}$, $a_{K\Lambda}$, $a_{\eta N}$. Once these constants are fixed to the $\pi N$ scattering data, the amplitudes involving vector mesons, as well as strangeness production according to $\pi N\to K\Lambda, K\Sigma$, can be predicted without introducing any new free parameters. The values of the subtraction constants can be found in Ref. \cite{Inoue:2001ip} or Sec. \ref{sec:read}. 

In this study we will make no attempt to introduce the $\pi\pi N$ channel or modify the interaction with a form factor as done in Ref. \cite{Inoue:2001ip}. This is because these modifications cannot easily be extended to the high energies of $\phi$ production. Also, the influence of the $\pi\pi N$ channel has been found rather small in isospin 1/2 (the reaction of interest, $\pi N\to\phi N$, is in pure isospin 1/2).

The amplitudes $T^{ij}$ from Eq.~(\ref{bse}) can be analytically continued to the complex plane of the scattering energy $s^{1/2}$. The amplitude has a pole on the second Riemann sheet that is identified with the resonance. 
Around the position of the pole at the complex value $M^*$, the amplitude can be expanded according to
\begin{equation}
   T^{ij}_{\rm pole} = \frac{g_ig_j}{s^{1/2}-M^*}\,.
   \label{poleapp}
\end{equation}
The expression in the numerator, which is given by the residue of the amplitude, determines the coupling strengths $g_i$ of the resonance to the different channels.

The pole positions of the resonance have been obtained in Ref. \cite{Jido:2007sm} at 
\begin{equation}
   M^* = 1537 -37\,i\;  \rm{MeV}  \label{eq:PPnstar}
\end{equation}
for the $n^{*}$ (neutral charge) and 
\begin{eqnarray}
   M^* = 1532 -37\,i\; \rm{MeV} \label{eq:PPpstar}
\end{eqnarray}
for the $p^{*}$ ($+1$ charge).
The values of the coupling constants $g_i$ are listed in Ref. \cite{Jido:2007sm}. It turns out that the $N^*(1535)$ couples strongly to $K\Lambda$ and $K\Sigma$ and thus has large strangeness components. 

We will refer to the expansion in Eq. (\ref{poleapp}) as {\it pole approximation} in the following. Close to the pole position, Eq. (\ref{poleapp}) will be a good approximation while further away the full amplitude might become quite different from a resonant shape. For example, even at $s^{1/2}=1535$ MeV the resonant shape of the $N^*(1535)$ of Eq. (\ref{poleapp}) is already modified by a considerable background in $\pi N$ scattering. 

Note that Eq. (\ref{poleapp}) provides only a first order approximation and does not take into account the energy dependence of the width. We will further discuss this issue in Secs. \ref{sec:cross} and \ref{sec:comments}.

\subsection{Vector meson couplings}
\label{sec:vemocoup}
\begin{figure}
\includegraphics[width=0.2\textwidth]{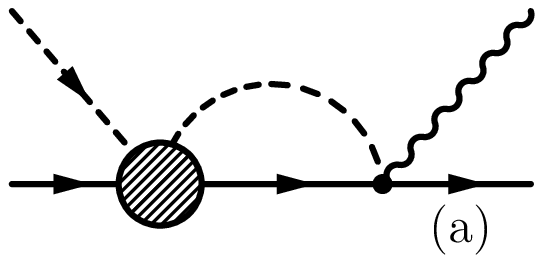}\hspace*{0.35cm}
\includegraphics[width=0.2\textwidth]{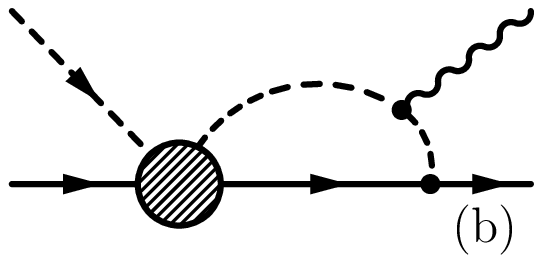}
\caption{Phi (wavy line) couplings to the meson baryon amplitude (shaded circle) via $K\Lambda$ and $K\Sigma$ loops (and $\pi N$ loops for the $\rho$). Left: Kroll-Ruderman term, right: meson pole term. The shaded circles represent the unitarized amplitude $\pi N\to KY,\,\pi N$.}
\label{fig:diagrams}
\end{figure}

In order to construct the transition $\pi N\to\phi N$, the $\phi$ is coupled to the kaon loops provided by the unitarized amplitude from Eq. (\ref{bse}). This is schematically shown in Fig. \ref{fig:diagrams}. For completeness, we will also consider the vector mesons $\omega$ and $\rho$.
The coupling of the vector mesons follows a similar scheme as in the case of photons, obeying similar constraints from gauge invariance as the photon. This requires, in principle, to couple the vector mesons to all possible meson and baryon propagators and vertices provided in the rescattering series from Eq. (\ref{bse}). In a recent study \cite{Jido:2007sm}, the electromagnetic form factors of the $N^*(1535)$ have been evaluated. There, some of the diagrams turn out to be of next-to-leading order. Only the diagrams shown in Fig. \ref{fig:diagrams} are of leading order. In practice, it is enough to calculate the meson pole term on the righthand side. Indeed, imposing gauge invariance, the contribution from the Kroll-Ruderman term can be taken into account without any explicit calculation. Then, the diagrams from Fig. \ref{fig:diagrams} form a gauge invariant subset of leading order diagrams. For a detailed discussion on gauge invariance and subleading terms, see Ref. \cite{Jido:2007sm}. The main point when adapting the formalism from photons to vector mesons is that the virtuality of the photon corresponds to the meson mass, $-Q^2=m_V^2$.

The transition amplitude, involving vector mesons, can be written as
\ba
T=T^{\mu\nu}\;\sigma_\mu\;\epsilon_\nu
\ea
where $\sigma^\mu\equiv(0,\vec{\sigma})$ is the spin operator of the baryon and $\epsilon$ the polarization vector of the vector meson (V). Lorentz invariance implies the amplitude to be of the general form  \cite{Marco:1999df,Roca:2006am,Doring:2007rz}
\ba
-i\,T^{\mu\nu}=a\,g^{\mu\nu}+b\,P^\mu P^\nu+c\,P^\mu k^\nu+d\,P^\nu k^\mu+e\,k^\mu k^\nu\nonumber \\
\label{genfo}
\ea
where $P$ is the total fourmomentum and $k$ is the momentum of the vector meson. Using the Lorentz condition $\epsilon^\mu\,k_\mu=0$ and the fact that  $\sigma^\mu\,P_\mu=0$ in the c.m. frame, the amplitude turns out to be
\ba
-i\,T=-a\;\vec{\sigma}\cdot\vec{\epsilon}-d\;\vec{\sigma}\cdot\vec{k}\;\epsilon^0\,P^0.
\label{tasde}
\ea
In the following, we neglect the second term which is proportional to the momentum of the vector meson $k$, i.e. small close to the threshold of vector meson production. Contracting the amplitude $T^{\mu\nu}$ with $k_\nu$, i.e. replacing $\epsilon_\nu\to k_\nu$, we obtain from the gauge invariance two conditions
for the coefficients $a$ to $e$ and in particular
\ba
a=-d\,P\cdot k-e\,k^2.
\label{acof}
\ea
Thus, the coefficient $a$ can be determined from $d$ and $e$. Evaluating $d$ and $e$ has the advantage that the contributions of loop amplitudes to these coefficients appear with two powers of loop momenta less than in case of $a$ as one directly see from Eq. (\ref{genfo}). In particular, the sum of the transition loops with $\phi$ couplings are finite and require no renormalization. 

In the present, nonrelativistic, formulation, the leading terms of the $1/M$ expansion are given by the two diagrams in Fig. \ref{fig:diagrams}. As the Kroll-Rudermann
term only contributes to $a$, it is enough to calculate $d$ and $e$ from the meson pole term and then use Eqs. (\ref{tasde}) and (\ref{acof}) in order to evaluate the
amplitude. The calculation of the meson pole term for finite $k$ is described in detail in Ref. \cite{Jido:2007sm}. Here, we only display the final result which is similar to the expressions in Ref. \cite{Jido:2007sm}. For the $MBB$ and the $VMM$ \cite{Palomar:2002hk,Klingl:1996by} vertices we have used the interaction Lagrangians
\ba
{\cal L}_{MBB} &=& 
 - \frac{D}{\sqrt 2 f} \, {\rm Tr} \left[ \bar B \gamma_{\mu} \gamma_{5} \{\partial^{\mu}\Phi,B\} \right]
 \nonumber \\ && 
 - \frac{F}{\sqrt 2 f}\, {\rm Tr} \left[ \bar B \gamma_{\mu} \gamma_{5} [\partial^{\mu}\Phi ,B] \right] \nonumber \\
{\cal L}_{VMM} &=& \frac{i\,G_V\,m_V}{\sqrt{2}\,f^2}\;{\rm Tr}(V^\mu\,[\partial_\mu P,P])
\label{ial} 
\ea
respectively. The matrices $V$ and $P$ represent the fields of vector and pseudoscalar mesons, respectively. An explicit expression for $V$ can be found in Ref. \cite{Palomar:2002hk}, the meson and baryon matrices $\Phi\equiv P$ and $B$ can be found in Ref. \cite{Jido:2007sm}.

The total amplitude for the process $\pi N\to \phi N$ is given by 
\ba
T_{\rm{tot}}^i=\sum_i T^{ij}a_j\;(\vec{\sigma}\cdot\vec{\epsilon})
\label{ttot}
\ea
with $T^{ij}$ being the unitarized $MB\to MB$ transition from Eq. (\ref{bse}) [$i=4$ for the initial state $\pi^-p$], $a_j$ is given by Eq. (\ref{acof}) and the sum is over all channels $j$ for the intermediate meson baryon state, i.e. the loop shown in Fig. \ref{fig:diagrams}. The coefficients $d$ and $e$ are given by
\ba
d_j&=&\frac{A_j\;g_A^j\;g_V}{16\pi^2}\int\limits_0^1 dx\;\int\limits_0^x dy\;\frac{2M_j(1-y)(1-x)}{S+i\epsilon},\nonumber\\
e_j&=&\frac{A_j\;g_A^j\;g_V}{16\pi^2}\int\limits_0^1 dx\;\int\limits_0^x dy\;\frac{M_j(1-y)(2y-1)}{S+i\epsilon}
\label{de}
\ea
where 
\ba
S&=&P^2x(1-x)+k^2y(1-y)-M_j^2(1-x)\nonumber \\
&-&m_j^2x-2P\cdot k(1-x)y.
\ea
The meson and baryon masses of the loop are given by $m_j$ and $M_j$. The axial coupling constants $g_A^j$ from the $MBB$ Lagrangian can be directly taken from Table III of Ref. \cite{Jido:2007sm}. The relevant coefficients from the $VMM$ couplings $A_j$ are $A_j^\phi=1/(\sqrt{2}\,f)$ and $A_j^\omega=-1/(2\,f)$ for $\phi$ and $\omega$ production, in all channels $j$ with kaons, and zero otherwise. For the $\rho^0$ case, $A^{\rho_0}=-1/(2\,f)$ for loops with $K^+$, $A^{\rho_0}=+1/(2\,f)$ for loops with $K^0$, $A^{\rho_0}=1/f$ for the loop with $\pi^-$, and  $A^{\rho_0}=-1/f$ for the loop with $\pi^+$. The coefficient $A^{\rho_0}=0$ for channels with $\pi^0$ and $\eta$. In Eq. (\ref{de}),
\ba
g_V=\frac{m_V\;G_V}{f^2}
\ea
where the vector coupling strength $G_V=56$ MeV, $m_V$ is the mass of the vector meson, and $f=93$ MeV.

So far, we have determined the amplitude for the reaction $\pi N\to\phi N$. We can also evaluate the effective coupling strengths $g_{\phi N}$, $g_{\omega N}$, and $g_{\rho^0 N}$ of the $N^*(1535)$ to the vector mesons. 
If in Eq. (\ref{ttot}) we substitute $T^{ij}$  by its pole approximation of Eq. (\ref{poleapp}) and $T_{\rm{tot}}^i$ by its pole approximation
\ba
T_{\rm{PA}}^i=\frac{g_i\;g_{VN^*}}{s^{1/2}-M^*}\,(\vec{\sigma}\cdot\vec{\epsilon}),
\label{phiprodpa}
\ea
we obtain
\ba
g_{VN^{*}}=\sum_j a_j\;g_j
\label{effve}
\ea
where the $g_j$ are the coupling strengths of the $N^*(1535)$ to the coupled channels of the model as determined from the residue of the pole according to Eq. (\ref{poleapp}). In Sec. \ref{sec:gi} we will show results for the effective couplings of vector mesons to the $N^*(1535)$.

Since the pole approximation is only good close to the pole, we should not use Eq. (\ref{phiprodpa}) as a substitute of Eq. (\ref{ttot}) which relies upon the full $T^{ij}$ amplitudes, quite different to their pole approximation when we move away from the pole as it is the case here.

In the derivation of the amplitude we have neglected the term with $d$ in Eq. (\ref{tasde}) that vanishes at the threshold of vector meson production. Consequently, we evaluate the intermediate loop at the threshold, $s^{1/2}=m_V+M$. Furthermore, we have $k^2=m_V^2$ and $P\cdot k=m_V(m_V+M)$. Then, the entire energy dependence for the reaction $\pi N\to\phi N$ comes from the unitarized amplitude from Eq. (\ref{bse}) or, in case of the pole approximation, from the denominator in Eq. (\ref{phiprodpa}). 

\section{Results}
\subsection{The couplings strengths of the $N^*(1535)$}
\label{sec:gi}
\begin{table}
\caption{The coupling of the $\phi$ to the $N^*(1535)$. For completeness, we also show the predictions for the $\rho N N^*(1535)$ and $\omega N N^*(1535)$ couplings.}
\begin{center}
\begin{tabular}{lll}
 \hline\hline
\hspace*{0.5cm}&
\multicolumn{2}{c}{
$g_{\rm{VB\to N^*(1535)}}$}\\
&this study&other studies\\
$\phi p$&\hspace*{0.25cm}$0.03-0.15\, i$\hspace*{0.2cm}&\\
$\phi n$&\hspace*{0.25cm}$0.04-0.17\, i$&\\
$\omega p$&$-0.03 + 0.15\, i$&\\
$\omega n$&$-0.03 + 0.16\, i$&\\
$\rho^0 p$&\hspace*{0.25cm}$0.63 - 0.04\, i$&$0.69$ to $0.89$ \footnote{extraction from PDB without $\rho$ form factor.} \cite{PDG}, $\pm 1.12$\footnote{in agreement with the PDB \cite{PDG}, implying a form factor for the offshell $\rho$.} \cite{Xie:2007qt} \\
$\rho^0 n$&$-0.64+0.04\,i$& \\ 
 \hline\hline
\end{tabular}
\end{center}
\label{tab:couplings}
\end{table}
The coupling strengths of the $N^*(1535)$ to $\phi N$, $\omega N$, and $\rho N$ are displayed in Table \ref{tab:couplings}. These values are valid at the respective vector meson production thresholds due to the approximations made in Sec. \ref{sec:vemocoup}. Yet, their size is a good approximation in the vicinity of the thresholds as the omitted term of the form $\epsilon^0\;\vec{\sigma} \cdot {\vec k}$ is small for small momenta ${\vec k}$ of the vector meson.
The calculation has been done in the particle base, which allows for isospin breaking from different masses. However, this effect is negligible and the values in Table \ref{tab:couplings} are almost isospin symmetric, i.e., same size and sign for the $\phi N$ and $\omega N$ couplings, $N=(p,n)$, and same size and opposite sign for the $\rho^0 N$ couplings.

The $\rho N$ coupling, predicted in the present study,
is in good agreement with the value extracted from the
PDB \cite{PDG} within experimental uncertainties. To show this
we write the width of the $N^*(1535)$ decaying into $\rho N$ in $s$-wave,
\ba
\Gamma_{N^*\to (N\rho[\pi\pi])_s}&=&\frac{3\,M_N}{M_{N^*}}\,\frac{g_{\rho^0p}^2\,f_\rho^2}{8\pi^3}\int\limits_{2m_\pi}^{M_{N^*}-M_N}dM_I\;p_N\tilde{k}\nonumber\\
&\times&
\frac{M_I^2-4m_\pi^2}{\left(M_I^2-m_\rho^2\right)^2+\left(M_I\,\Gamma_\rho\right)^2}
\ea
where 
\ba
p_N=\frac{\lambda^{1/2}(M_{N^*}^2,\,M_N^2,\,M_I^2)}{2\,M_{N^*}}
\ea
and $\tilde{k}=(M_I^2-4m_\pi^2)^{1/2}/2$
and $f_\rho$ is the coupling of the $\rho^0$ to two pions, $f_\rho=6.02$, while 
$g_{\rho^0p}$ is the coupling ($g_{\rho^0p}\, \vec{\sigma}\cdot\vec{\epsilon}$) of the $N^*(1535)$ to $\rho^0 p$.

Taking the branching ratio of 0.02 quoted in Ref. \cite{PDG} and the width
$\Gamma_{N^*}=150$ MeV we obtain $g_{\rho N}=0.89$, and if we take  $\Gamma_{N^*}=90$ MeV
we obtain $g_{\rho N}=0.69$.  Given the large span of the branching ratio quoted in Ref.
\cite{PDG}, the result that we obtain is consistent with present 
experimental
data on this magnitude. This is a good signal that our predicted couplings
are realistic.

The coupling of $1.12$ reported in Ref. \cite{Xie:2007qt} is equivalent to ours because in Ref. \cite{Xie:2007qt}
an extra form factor for the off shell $\rho$ is implemented, whose omission leads to the same value of $g_{\rho N}=0.89$ found here. We do not have explicitly this form factor since the loop calculations
incorporate automatically any off shell dependence. In our case the dependence
of this coupling on the mass of the $\rho$ is moderate, and changes of the mass
by about 50 MeV do not change appreciably the coupling. Even going to a value of
$s^{1/2}= 1535$ MeV there is less than 20 \% decrease in the modulus of the value of the
coupling reported in Table \ref{tab:couplings}.

The $\rho$ couples to both the $\pi N$ and the $KY$ channels of the $N^*(1535)$ via the loops shown in Fig. \ref{fig:diagrams}. Omitting the $KY$ loops, the modulus of the $\rho NN^*(1535)$ coupling constant decreases by 40 \% and is not in agreement with the PDG value any more. This shows that the strangeness channels in the $N^*(1535)$ play an important role. 

The couplings to $\omega N$ and $\phi N$ are smaller than $g_{\rho N}$. The reason is, in case of the $\phi N$ and $\omega N$ couplings, that for the $K\Lambda$ intermediate state the combination of the $N^*K\Lambda$ vertex, together with the $\bar{K}K\phi$ and $K\Lambda N$ vertices, almost cancels the corresponding combination from the $K\Sigma$ intermediate loop. This is not the case for the coupling of the $N^*(1535)$ to $\rho N$. See also Sec. \ref{sec:comments} for an estimate of the theoretical errors of $g_{\phi N}$.  

At the higher energies of $\phi$ production, the picture completely changes, because one should rely on the full, and energy dependent, amplitude $MB \to MB$ rather than on the couplings $g_{KY}$ that are extracted at the pole of the $N^*(1535)$. At the high energies of $\phi$ production, the contributions from the $K\Lambda$ and $K\Sigma$ channel no longer cancel, and the cross section from the full amplitude is much larger than the one from the pole approximation, which relies on the couplings extracted at the pole position. This issue will be further discussed in the following subsections.

The couplings of the $N^*(1535)$ to $KY$, $\pi N$, and $\eta N$ can be found in Ref. \cite{Jido:2007sm}. Together with the values from Table \ref{tab:couplings}, we can summarize some properties of the $N^*(1535)$ in a schematic ordering of its coupling strengths according to $g_{K\Sigma},g_{K\Lambda},g_{\eta N}>g_{\pi N},g_{\rho N}>g_{\omega N},g_{\phi N}$. While the large $KY, \eta N$ couplings are responsible for the formation of the dynamically generated pole, we have found that the coupling to $\rho N$ is of the size of $\pi N$, i.e. smaller. This also means that including the $\rho N$ channel in the coupled channel approach would not change much the properties of the $N^*(1535)$ and the channel space used in Ref. \cite{Inoue:2001ip} is sufficiently large.

The $\omega N$ coupling to the $N^*(1535)$ in Table \ref{tab:couplings} is quite small. It has been shown recently \cite{Nakayama:2008tg} that it is difficult to extract this coupling phenomenologically in an unambiguous way.

\subsection{$\phi$ and strangeness production}
\label{sec:cross}
\begin{figure}
\includegraphics[width=0.45\textwidth]{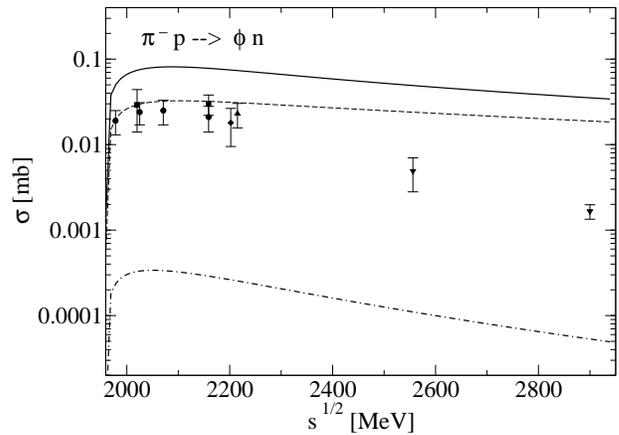}
\caption{Reaction $\pi^-p\to\phi n$. The solid line shows the result using the full, energy dependent amplitude, while the dashed-dotted line shows the pole approximation. The dashed line shows the result using the full amplitude after a readjustment of the subtraction constants, as described in Sec. \ref{sec:read}. The complete data reference can be found in Ref. \cite{Landolt}.}
\label{fig:pimp_to_phin}
\end{figure}

\begin{figure}
\includegraphics[width=0.45\textwidth]{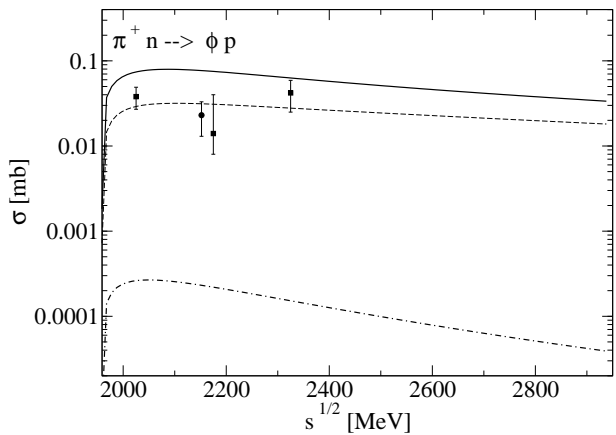}
\caption{Reaction $\pi^+n\to\phi p$. Theoretical curves as described in Fig. \ref{fig:pimp_to_phin}. Complete data reference in Ref. \cite{Landolt}, except the squares which are from Ref. \cite{Davies:1970ex}.}
\label{fig:pipn_to_phip}
\end{figure}

The $\phi$ production is shown in Figs. \ref{fig:pimp_to_phin} and \ref{fig:pipn_to_phip}. The process $\pi N\to \phi N$ is in pure isospin 1/2 and we would expect very similar results for $\pi^-p\to\phi n$ and $\pi^+n\to\phi p$. This is indeed the case both for the data and the theoretical curves as Figs. \ref{fig:pimp_to_phin} and \ref{fig:pipn_to_phip} show.

The solid lines show the results using the full solution of the Bethe-Salpeter equation (\ref{bse}) for the $T^{ij}$ amplitudes in Eq. ({\ref{ttot}}), while the dashed-dotted lines represent the $N^*(1535)$ pole approximation of the $MB\to MB$ transition from Eqs. (\ref{poleapp},\ref{phiprodpa}). The dashed line shows the outcome after a readjustment of the parameters discussed in Sec. \ref{sec:read}.

The smallness of the result when using the pole approximation is consistent with the small value that we obtain for the coupling in Table \ref{tab:couplings}, $|g_{\phi N}|=0.17$. As a reference, this value is about one order of magnitude below the value determined in a fit to $\phi N$ data in Ref. \cite{Xie:2007qt}, $g_{\phi N}=1.2$ although our pole approximation and the approach of Ref. \cite{Xie:2007qt} are rather different as we explain below. However, this gives us a rough idea of values needed to get a fit to the $\pi N\to \phi N$ data assuming it provided by the $N^*(1535)$ resonance alone.

If we use the full solution of the Bethe-Salpeter equation (\ref{bse}) instead of the pole approximation from Eq. (\ref{phiprodpa}), we obtain the solid lines in Figs. \ref{fig:pimp_to_phin} and \ref{fig:pipn_to_phip}. The result is much larger due to the following reasons: First, the form of the pole approximation from Eq. (\ref{poleapp}) makes the contribution automatically small 400 MeV above the nominal mass of the $N^*(1535)$. Second, the smallness of $g_{\phi N N^*}$ has been traced back to a cancellation of $K\Lambda$ and $K\Sigma$ intermediate loops as discussed in Sec. \ref{sec:gi}. In the full, energy dependent amplitude, this cancellation is no longer valid at $s^{1/2}\sim 2$ GeV. As a consequence, the pole of the $N^*(1535)$ plays a minor role compared to the full amplitude which delivers a solution much closer to the $\phi N$ production data.

What we learn from there is that the pole approximation of Eq. (\ref{phiprodpa}) for the $\pi N\to\phi N$ reaction is quite bad, as a consequence of which the concept of the $\phi N^*$ coupling that we obtain is not very useful. In Ref. \cite{Xie:2007qt} an empirical amplitude mediated by the $N^*(1535)$ excitation is used, with a Breit-Wigner form of the type of Eq. (\ref{phiprodpa}), but incorporating the energy dependent width and an extra form factor. In Ref. \cite{Xie:2007qt} several options were studied, including the possibility of having contributions from other resonances, like the $N^*(1650)$, $N^*(1710)$, $N^*(1720)$, $N^*(1900)$ but using different arguments the dominance of the $N^*(1535)$ was suggested and the model based upon single $N^*(1535)$ excitation was then used to study the $pp\to pp\phi$ reaction. The coupling $g_{\phi N^*}$ of the amplitude of Ref. \cite{Xie:2007qt} is fitted to reproduce the $\pi N\to\phi N$ data. This is a parametrization of some data with a specific form of an amplitude which is different to our pole approximation of Eq. (\ref{phiprodpa}) or the full amplitude of Eq. (\ref{ttot}). Hence the comparison of the coupling $g_{\phi N^*}$ obtained in Ref. \cite{Xie:2007qt} and the one found here would be improper. However, even with the differences in the amplitudes, one can qualitatively understand, from the results in Figs. \ref{fig:pimp_to_phin} and \ref{fig:pipn_to_phip} with the simple pole approximation, why a larger $\phi N^*$ coupling is needed in Ref. \cite{Xie:2007qt} to reproduce empirically the $\pi N\to\phi N$ data.

The important point, that we should stress here, is that the chiral unitary approach, adjusting only a few subtraction constants around the $N^*(1535)$ energy region to fit the $\pi N\to\pi N$ data, is able to make a prediction for the $\pi N\to\phi N$ cross section close to $\phi N$ threshold, within a factor of two, without the need to fit any extra parameters. The model of Ref. \cite{Xie:2007qt} is a parametrization of the data of $\pi N\to\phi N$ assuming $N^*(1535)$ dominance. 
It is an effective parametrization of the full $\pi N \to \phi N$ amplitude with a single resonance. However, in the present study, although our full amplitude has only one pole corresponding to the $N^*(1535)$, it also contains a large non-resonant background contribution. The value of the work in Ref. \cite{Xie:2007qt} can be seen from another perspective: once an empirical parametrization of the $\pi N\to\phi N$ data is done, such information can be used in related processes like the $pp\to pp\phi$ reaction and, indeed, it is shown in Ref. \cite{Xie:2007qt} that the cross sections of both reactions can be reproduced simultaneously.

At this point it is illuminating to consider the $\omega$ production according
to $\pi N\to\omega N$ (not shown here but calculated). The cross section is of
similar size and has a similar energy dependence as in $\pi N\to\phi N$, while
the data in $\omega$ production reaches $\sigma=1.5$ mb already 50 MeV above
the $\omega N$ threshold. Thus, the present model, using the full amplitude, is
more than a factor of ten below data. This illustrates that the present model
is indeed suited for the OZI-violating $\phi$ production, where it matches the
data much better; for $\omega$  production, in contrast, resonances and their
OZI-allowed couplings to $\omega N$ will dominate the cross section, and the
present model delivers only a small part of the amplitude.

\begin{figure}
\includegraphics[width=0.45\textwidth]{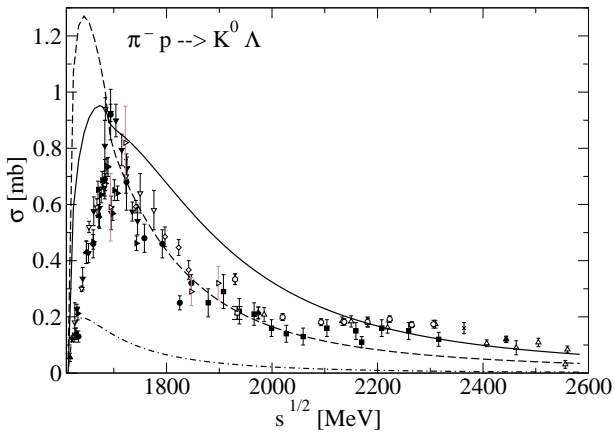}
\caption{Reaction $\pi^-p\to K^0\Lambda$. Theoretical curves as described in Fig. \ref{fig:pimp_to_phin}. Complete data reference in Ref. \cite{Landolt}.}
\label{fig:pimp_to_KzL}
\end{figure}

\begin{figure}
\includegraphics[width=0.45\textwidth]{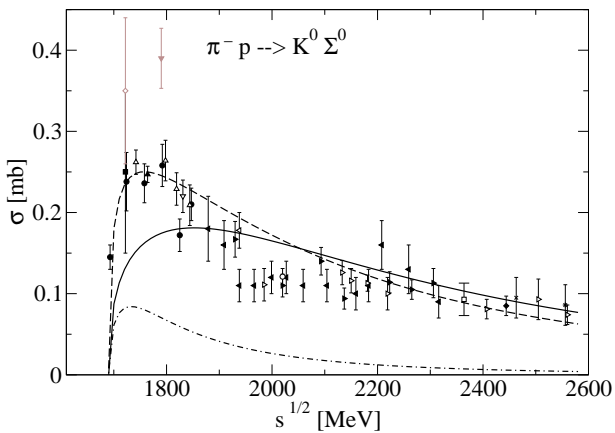}
\caption{Reaction $\pi^-p\to K^0\Sigma^0$. Theoretical curves as described in Fig. \ref{fig:pimp_to_phin}. Complete data reference in Ref. \cite{Landolt}.}
\label{fig:pimp_to_KzSz}
\end{figure}

\begin{figure}
\includegraphics[width=0.45\textwidth]{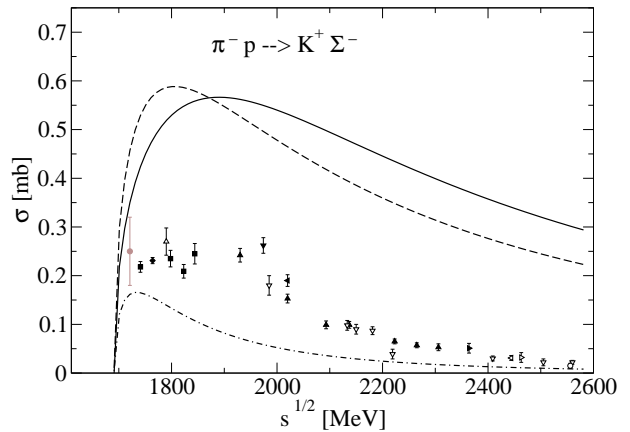}
\caption{Reaction $\pi^-p\to K^+\Sigma^-$. Theoretical curves as described in Fig. \ref{fig:pimp_to_phin}. Complete data reference in Ref. \cite{Landolt}.}
\label{fig:pimp_to_KpSm}
\end{figure}

The solution using the full amplitude, shown with the solid lines in Figs. \ref{fig:pimp_to_phin} and \ref{fig:pipn_to_phip}, still overestimates the data by a factor of around 2.5.
As described in Sec. \ref{sec:model}, the full amplitude of $MB\to MB$ has been finetuned to fit $\pi N$ scattering data from the $\pi N$ threshold up to $s^{1/2}\sim 1.6$ GeV. Thus, we cannot expect a precise prediction at the high energies of $\phi$ production. Yet, we can estimate the expected precision by investigating strangeness production within the present model: The $\phi$ couples to the meson baryon amplitude through its decay channel into $\bar{K}K$ as described in Sec. \ref{sec:model}. Strangeness production according to $\pi N\to KY$, where $Y=\Lambda,\, \Sigma$ can be evaluated from the present model by choosing the corresponding final state $j$ in the transition amplitude $T^{ij}$ from Eq. (\ref{bse}). The cross sections for $\pi N\to YK$ are shown in Figs. \ref{fig:pimp_to_KzL} to \ref{fig:pimp_to_KpSm}.

Again, the solid lines show the result using the full amplitude, the dashed-dotted lines show the pole approximation and the dashed lines represent the result after a readjustment of the parameters as discussed in Sec. \ref{sec:read}. The pole approximation is quite different from the outcome with the full amplitude and lies much below data for all reactions. This becomes worse and worse at energies further away from the nominal $N^*(1535)$ mass as one would expect. In particular, the $\pi^-p\to K^0\Lambda$ reaction shows that the opening of the $K\Sigma$ channel plays an important role. This structure is contained in the full amplitude, while in the pole approximation from Eq. (\ref{poleapp}), only the information about the sub-threshold $K\Lambda$ and $K\Sigma$ amplitudes is present. These findings imply also a caveat for effective Lagrangian approaches using sub-threshold resonances: we have seen that the contribution from the pole can become very small, and thus a fit to data, using only resonance contributions, can easily lead to an overestimation of the coupling strength to the sub-threshold resonance.

In the following, we concentrate on the full amplitude, shown with the solid lines.
For the reactions $\pi^-p\to K^0\Lambda$ and $\pi^-p\to K^0\Sigma^0$ we observe fair agreement in the cross sections, even in the region of $\phi$ production around  $s^{1/2}\sim 2$ GeV. This behavior is comparable to previous findings for a similar model \cite{Nacher:1999vg}. In Ref. \cite{Mosel} the bump at $s^{1/2}=1.7$ GeV in $\pi^-p\to K^0\Lambda$ is interpreted as an interplay between the $P_{13}(1720)$ resonance and the opening of the $K\Sigma$ and $\omega N$ channels, although the bump can be also well described by a pure $P_{11}(1710)$ contribution in a reduced channel space. The picture changes again once photoproduction data are included \cite{Mosel}. The quality of the differential cross section data is not good and it is difficult to pin down the contribution from different channels and partial waves uniquely. In the present model, the structure at  $s^{1/2}=1.7$ GeV is given entirely by the opening of the $K\Sigma$ channel within the coupled channel dynamics in pure $s$-wave. 

At the right shoulder of the bump there is an overprediction of the data, that reaches up to $s^{1/2}\sim 2$ GeV, i.e. the region of $\phi$ production. Yet, the overall shape is fairly well reproduced, and it is noteworthy, that the strength and energy dependence of the cross sections in $\pi^-p\to K^0\Lambda$ and $\pi^-p\to K^0\Sigma^0$ are predictions and involve no free parameters except those previously fixed in $\pi N$ scattering [see Sec. \ref{sec:model}]. 

For the $\pi^-p\to K^0\Sigma^0$ reaction shown in Fig. \ref{fig:pimp_to_KzSz}, the present calculation (solid line) leaves room for additional structures at around $s^{1/2}=1.75$ GeV. There, the excess of the cross section could indicate a resonant contribution. In Ref. \cite{Mosel}, a sizable contribution from the $P_{11}(1710)$ resonance in $J^P=1/2^+$ is found. In the same reference, a strong $s$-wave contribution close to threshold has been found which supports the present $s$-wave calculation.

For the reaction $\pi^-p\to K^+\Sigma^-$ shown in Fig. \ref{fig:pimp_to_KpSm}, the present study predicts a cross section that is too large compared to data. Close to threshold the model deviates by a factor of around two, while at higher energies the discrepancy is even larger. At $s^{1/2}\sim 2$ GeV, which is the energy where $\phi$ production starts, the model overpredicts the cross section by a factor of around 3. 

We cannot expect a better precision for $\phi$ production than the precision in pion-induced strangeness production, and from the $\pi^-p\to K^0\Lambda$ and
$\pi^-p\to K^+\Sigma^-$ reaction we have learned that a factor of 2 to 3 deviation must be admitted. This is indeed what we found for the $\phi$ production.

\begin{figure}
\includegraphics[width=0.45\textwidth]{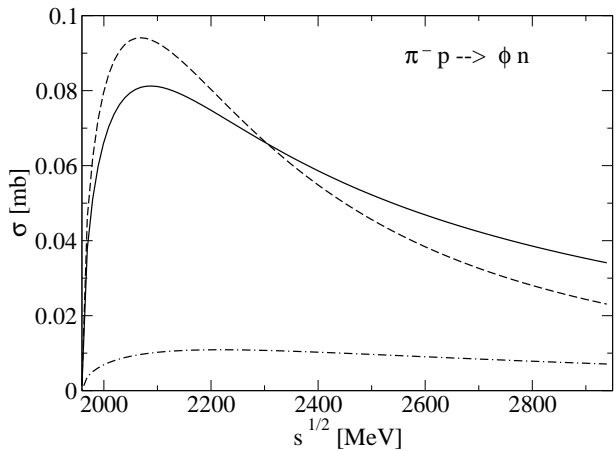}
\caption{Individual contributions of intermediate states to the $\pi^-p\to\phi n$ cross section. Dashed line: $K\Lambda$ intermediate state. Dashed dotted line: $K\Sigma$ intermediate state. Solid line: Sum of the contributions.}
\label{fig:pimp_to_phin_indi}
\end{figure}

It is instructive to see the different weights of terms in the $\phi$ production. The intermediate states with $\phi$ couplings [see Fig. \ref{fig:diagrams}] in the reaction $\pi^-p\to\phi n$ are $K^+\Sigma^-$, $K^0\Sigma^0$, and $K^0\Lambda$. In Fig. \ref{fig:pimp_to_phin_indi} we show the contributions from these intermediate states.
The contribution from the $K^0\Lambda$ state largely dominates. Indeed, setting this loop to zero, the $\phi$ production cross section is reduced by a factor of around ten, i.e. a factor of three below data. The reason for the large contribution can be found in the $NK\Lambda$ vertex, which is around three times larger than the $NK\Sigma$ vertex. Additionally, the $K\Lambda$ threshold is lower than the $K\Sigma$ one and the intermediate loop function is larger for the $K\Lambda$ state. The dominance of the $K^0\Lambda$ state also means that the observed discrepancies in the reaction $\pi^-p\to K^+\Sigma^-$ [see Fig. \ref{fig:pimp_to_KpSm}] will have only moderate influence in the $\pi^-p\to\phi n$ reaction. The overprediction for the cross section in $\pi^-p\to\phi n$ is rather tied to the overprediction  of the data in $\pi^-p\to K^0\Lambda$ at around $s^{1/2}\sim 2$ GeV [see Fig. \ref{fig:pimp_to_KzL}]. Indeed, in Sec. \ref{sec:read} we will see that a better description of the  $\pi^-p\to K^0\Lambda$ cross section leads to a better description of the $\pi N\to\phi N$ reactions.

Note that the dominance of the $K^0\Lambda$ intermediate state is quite different to the cancellation pattern between the $K\Lambda$ and $K\Sigma$ intermediate states which we have observed in the determination of the coupling strengths of the $N^*(1535)$ to $\phi N$ in Sec. \ref{sec:gi}. Again, this is due to the fact that the full, energy dependent $MB\to MB$ amplitude contains much more information than the pole approximation extrapolated to the high energies of $\phi$ production. It is then the full amplitude which delivers a much more realistic description of the data than the pole approximation. 

\subsection{Readjusting the parameters of the amplitude}
\label{sec:read}
The only free parameters in the present study are the four subtraction constants of the loop function from Eq. (\ref{eq:gpropdr}), which have been fitted to $\pi N$ scattering data in $S_{11}$ and $S_{31}$ from threshold up to $s^{1/2}\sim 1.6$ GeV. Varying the subtraction constants around their ''natural'' value \cite{Hyodo:2008xr} is a common way to absorb effects of different nature which are not explicitly included in the strong $MB\to MB$ amplitude. These can be effects of higher order Lagrangians, which can also be related to genuine resonance poles \cite{Hyodo:2008xr}. The role of the subtraction constants in unitarized chiral perturbation theory has been recently discussed in Ref. \cite{Hyodo:2008xr}. Also, there are higher order relativistic effects which can be absorbed in the values of the subtraction constants [see a discussion in Ref. \cite{Jido:2007sm}]. Many of these higher order corrections are small or at least slowly varying with energy so that a constant in energy is sufficient to absorb them. However, in the present study we are interested in an energy region 500 MeV above the range where the subtraction constants have been fixed originally \cite{Inoue:2001ip}. Thus, the energy dependence of those effects, that were absorbed in the subtraction constants, might require a slight readjustment. 

Indeed, as Fig. \ref{fig:pimp_to_KzL} shows, the $\pi^- p\to K^0\Lambda$ cross section is too large compared to data at $s^{1/2}=2$ GeV, about a factor of two. When readjusting the subtraction constants in Eq. (\ref{eq:gpropdr}) we require that the data in $\pi^- p\to K^0\Lambda$ is matched at $s^{1/2}=2$ GeV. At the same time, we require that the cross sections stay similar in the other $\pi N\to KY$ reactions studied. Additionally, we require that the pole of the $N^*(1535)$ remains, although it might move slightly. Finally, we require that the set of subtraction constants is as close as possible to the original values from Ref. \cite{Inoue:2001ip}. 

A readjustment which fulfills all these requirements is given with $a_{K\Sigma}=-2.8$, $a_{K\Lambda}=2.6$, $a_{\pi N}=3.6$, $a_{\eta N}=0.6$ (compared to the respective values of $-2.8$, $1.6$, $2.0$, $0.2$ in the original fit from Ref. \cite{Inoue:2001ip}). The resulting cross sections are plotted in Figs. \ref{fig:pimp_to_phin} to \ref{fig:pimp_to_KpSm} with the dashed lines. 

In order to show the quality of the new fit we quote the values of the $\chi^2$ for the $\pi^-p\to K^0\Lambda$ reaction for the energy region from $1875$ MeV till $2125$ MeV: the $\chi^2$ per degree of freedom is $\chi/{\rm D.o.f.}= 38$ with the original set of subtraction constants, while after the readjustment, $\chi/{\rm D.o.f.}=7.9$, which is still a high value. However, as Fig. \ref{fig:pimp_to_KzL} shows, the experimental data from Ref. \cite{schwartzthesis} (open circles) seem to be incompatible with the rest of the data from Ref. \cite{Landolt} in that energy region. Indeed, omitting the three data points from Ref. \cite{schwartzthesis} in the fitted region, $\chi/{\rm D.o.f.}=1.1$, which is a clear signal of incompatible data. 

The reduction of the cross section in $\pi^-p\to K^0\Lambda$ obtained in the new fit leads to a reduced cross section in the $\phi$ production reactions, because the $K^0\Lambda$ intermediate state dominates as discussed at the end of Sec. \ref{sec:cross}. With the readjusted subtraction constants, we obtain a fairly good description of the data. We would not claim to be able to describe the higher energy data in $\phi N$ production at $2.6$ and $2.9$ GeV in Fig.  \ref{fig:pimp_to_phin} because these energies are much too high for the present model.

\section{Overview and Discussion }
\label{sec:comments}
Along the paper we have clearly stated the approximations done and the level of
accuracy expected. We found the $N^* \phi N$ coupling very small because of
strong cancellations between the intermediate $K \Lambda$ and $K \Sigma$
channels. Since the description of the $\pi^- p$ to $K \Lambda$ and $K \Sigma$
cross sections was only qualitative, we can infer some uncertainty in the $N^*$
couplings to $K \Lambda$ and $ K \Sigma$, although we should stress that the results
of the chiral unitary approach are much better at energies around the the
$N^*(1535)$ where the model has been fitted to the data of $\pi N$ scattering.
We can get an idea of these uncertainties by recalculating the 
$N^* \to K \Lambda$ and $N^* \to K \Sigma$ couplings with the new set of
parameters from Sec. \ref{sec:read}, and from this the new $N^* \phi N$ coupling. The results can be seen in
Table \ref{tab:new_coupl}. The coupling constants have been extracted through an expansion of the amplitude around the pole as discussed following Eq. (\ref{poleapp}).

\begin{table}
\caption{Couplings of the $N^*(1535)$, resulting after a readjustment of the subtraction constants [Sec. \ref{sec:read}]. The values in brackets show the size of the original couplings, corresponding to the subtraction constants from Ref. \cite{Inoue:2001ip}. The first three couplings correspond to the scattering problem of Ref. \cite{Inoue:2001ip}. The last three are obtained within the formalism of the present paper.}
\begin{center}
\begin{tabular}{lll}
 \hline\hline
\hspace*{1cm}&
\multicolumn{2}{c}{
$g_{\rm{N^*(1535)}}$}\\
&After Readjustment\hspace*{0.3cm}&original\\
$K^+\Sigma^-$&\hspace*{0.25cm}$2.09 + 0.17\, i$&(\hspace*{0.25cm}$2.20-0.17\, i)$\\
$K^0\Sigma^0$&$-1.49 - 0.13\, i$&$(-1.56+0.12 \, i)$\\
$K^0\Lambda$&\hspace*{0.25cm}$1.79 + 0.26\, i$&(\hspace*{0.25cm}$1.39-0.08\, i)$\\
$\phi n$&\hspace*{0.25cm}$0.03-0.41\,i$&(\hspace*{0.25cm}$0.04-0.17\, i)$\\
$\omega n$&\hspace*{0.25cm}$0.02+0.28\,i$&$(-0.03 + 0.16\, i)$\\
$\rho^0 n$&$-0.66-0.13\,i$&$(-0.64+0.04\,i)$ \\ 
 \hline\hline
\end{tabular}
\end{center}
\label{tab:new_coupl}
\end{table}

We see that the $N^* K \Lambda$ and $N^* K \Sigma$ couplings are rather stable
but, due to the cancellations mentioned, the $N^* \phi N$ coupling changes more
drastically, by about a factor 2.5 in modulus. With the new $N^* \phi N$ 
coupling the $\pi N \to \phi N$ cross section of Figs. \ref{fig:pimp_to_phin} and \ref{fig:pipn_to_phip} in the pole 
approximation would be increased by about a factor of five. Yet, the discrepancies
in about two orders of magnitude with the data remain.

  Once again, the results come to reinforce our comments in Sec. \ref{sec:cross} that the pole approximation from Eq. (\ref{phiprodpa}) is quite bad and, as a consequence, the concept of the $\phi N^*$ coupling associated to that approximation is not very useful.
  
  We would like to come back to the issue of the meaning of the $N^*(1535)$ as a
dynamically generated resonance. The approach of Ref. \cite{Inoue:2001ip} relies upon the use of the
lowest order chiral Lagrangians to construct the kernel of the interaction and
subtraction constants which are fitted to the $\pi N$ data. Unlike the case of
the $\Lambda(1405)$, where all the channels require the same subtraction
constant \cite{bennhold}, or equivalently, a unique cut off in all channels 
to regularize the loop functions \cite{ramos}, the case of the $N^*(1535)$
requires different ones in different channels, yet of natural size. This was
interpreted in Ref. \cite{Hyodo:2008xr} as an indication of the presence of extra non meson baryon
components, presumably $3q$ states. The flexibility on the choice of subtraction
constants, which regularize the loop functions, allows one to take into account
phenomenologically  such extra components in the scattering problem.

However, this is no longer the case in the radiative decay $N^* \to N \gamma$
where the loops with photon couplings are proved finite \cite{Jido:2007sm} and there is no freedom to fit anything.
Deficiencies in the theoretical framework will be reflected there in the
inability to describe these helicity amplitudes. In this sense it is worth
noting that these amplitudes are fairly well described by the couplings provided
by the chiral unitary approach \cite{Jido:2007sm}. Some discrepancies arise in the form
factors around  $Q^2=1\, $GeV$^2$ where the theoretical form factor falls too fast
and a compact three quark component like in the chiral quark models would
definitely help in producing a slower fall down \cite{thomas,tegen}.  The
acceptable results for the helicity amplitudes at $Q^2=0$, compared with the most
recent determination of the MAID2007 analysis \cite{maid}, indicate that the
baryon meson components are still the dominant ones in the $N^*(1535)$ wave
function and the relatively stable couplings of the $N^*$ to the meson baryon
components provided by the chiral unitary approach \cite{Inoue:2001ip} are realistic. Let us
also mention that the discrepancies with the data in the $Q^2$ dependence around $Q^2=1\, $GeV$^2$
are of the order of 20 \% for amplitudes normalized equally at $Q^2=0$, an
amount still lower than the uncertainties accepted here in the determination of
the $\pi N \to \phi N$ cross section.

Finally we would like to comment about another approach to the $\pi^-p\to\phi n$ reaction, followed in Refs. \cite{Nakayama:1999jx,Titov:2000bn,Titov:2001yw} in which the leading process is $t$ channel $\rho$ exchange relying upon the $\phi\to\rho\pi$ anomalous decay and form factors which are fitted to the experiment. A good description is obtained with this approach but in the absence of form factors the use of the $\rho$ exchange amplitude overshoots the cross section by about a factor of ten. The results of the present paper, where with no fits one already obtains the $\pi^-p\to\phi n$ cross section at the qualitative level, forces one to reopen this issue. A possibility is to have an OZI violating contact term for the $\pi^-p\to\phi n$ reaction. We can also think of additional terms with the same topology as the $t$ channel $\rho$ exchange. Particularly, a term where one has the $\pi\phi b_1(1235)$ vertex and the $b_1$ is exchanged and coupled to the nucleon. In recent theories where the axial vector meson  are dynamically generated from the interaction of vector mesons and pseudoscalars \cite{Lutz:2003fm,Roca:2005nm,Geng:2006yb} one finds the $b_1$ from the interaction of coupled channels namely $K\bar{K^*}$, $K^*\bar{K}$, $\phi\pi$. The coupling of the $b_1$ to $\phi\pi$ is found sizeable in \cite{Roca:2005nm}. Unfortunately, although some studies are devoted to the coupling of the $a_1$ to the nucleon \cite{Gasparyan:2003fp}, little is known about the coupling of $b_1$ to the nucleons, which does not allow us to proceed further in an evaluation of the actual contribution of this new term. However, the argument clearly indicates that the issue of the contribution of $t$ channel exchange is not settled with just the $\rho$ exchange. The findings of the present paper could stimulate work in this direction.

\section{Conclusions}
The study done here has allowed us to draw interesting conclusions. The
simultaneous study of the  $\pi^- p \to \phi n$ and the
$\pi^- p \to K^0 \Lambda$, $\pi^- p \to K^0 \Sigma ^0$ and $\pi^- p \to K^+ \Sigma ^-$
 reactions allowed us to get an idea of the 
level of
accuracy of the model used, from the comparison of the theoretical 
results with
the data for the $\pi^- p \to K^0 \Lambda$, $\pi^- p \to K^0 \Sigma 
^0$ and $\pi^- p \to K^+ \Sigma ^-$
 reactions. In the worse of the cases we found
discrepancies of about a factor 2.5.  We should not expect hence a better
agreement with data than such a factor for the $\pi^- p \to \phi n$ reaction
which occurs at higher energies from the $N^*(1535)$. The first finding 
of the
study was that the $N^*(1535)$ has indeed a non negligible coupling to
the $\phi N$ state. The second
finding was that using the simple pole approximation for the $\pi N \to \phi N$ 
reaction with the $N^* \phi N$ coupling found, we
obtain a too low cross section compared with experiment. 

We discussed that the simple pole approximation is very bad and one should rather use the full amplitude for the $\pi N\to\phi N$ reaction obtained by means of the full $\pi N\to K\Lambda,\,K\Sigma$ amplitudes.
The use of these full amplitudes
instead of their simple pole approximation 
gives rise to
a $\pi^- p \to \phi n$ cross section even larger than experiment but not
much different than
the difference seen in the $\pi^- p \to K^0 \Lambda$ reaction within the 
same
model. The consequences one draws from these results is that,
 when we study a region so far away from the pole, the
whole amplitudes $\pi^- p \to K^+ \Sigma ^-$, $\pi^- p \to K^0 \Sigma 
^0$ and
$\pi^- p \to K^0 \Lambda$ should be used rather than their pole 
approximation.

We also observed that the $\pi^- N \to \phi N$ reaction was dominated by the
intermediate $K^0 \Lambda$ state. Then, using some freedom in the model
for $MB \to MB$ which
allowed us to change the subtraction constants moderately, we could obtain a
better cross section for the $\pi^- N \to  K^0 \Lambda$ reaction around the
energy of threshold $\phi N$ production. By using this new input in the
$MB \to MB$ model then we obtained good cross sections for the
$\pi^- N \to \phi N$ reaction, within present experimental errors which 
are as
large as a factor of two.

 Should the experimental data improve in the future, further refinements 
would
be needed in the work done here. Indeed, in Ref. \cite{Jido:2007sm} it was shown 
that for large 
$Q^2$ values of the order
of $1$~GeV$^2$, as needed here, relativistic corrections could induce 
changes of
the order of 30-40 \%. Furthermore, we should
take into account that the coupling of $\phi N$ to $\pi N$, apart from
the source studied by us, could
also get contribution from a direct coupling if a more sophisticated 
approach to
the $\pi N$ reactions would be  followed which would include vector baryon
components in addition to the pseudoscalar baryon ones of
 \cite{Kaiser:1995cy,Nacher:1999vg,Nieves:2001wt,Inoue:2001ip,Hyodo:2002pk}.
This enterprise would be possible by extending to SU(6) the work based 
on SU(3)
 of the former references. The formalism for this extension has been 
carried
 out in \cite{GarciaRecio:2005hy} although not applied to the problem we are
 dealing with here. Our study could stimulate further work along this
 direction.

\section{Acknowledgments}
We would like to thank J. Haidenbauer for useful comments.
This work is partly supported by
DGICYT contract number FIS2006-03438,
and the National Natural Science Foundation
of China and the Chinese Academy of Sciences under
project number KJCX3-SYW-N2.
This research is  part of
the EU Integrated Infrastructure Initiative  Hadron Physics Project under
contract number RII3-CT-2004-506078. This study is also supported by a grant of the DFG (Deutsche Forschungsgemeinschaft).

\end{document}